\documentstyle[11pt]{article}
\begin{document}
\begin{titlepage}
\begin{flushright}
CfPA 98-th-05\\
astro-ph/xxyyzzz\\
(Submitted to {\bf Physical Review})\\
\end{flushright}
\begin{center}
\Large
{\bf A Fast Method For Bounding The CMB Power Spectrum Likelihood Function}\\
\vspace{1cm}
\normalsize
\large{Julian Borrill}\\
\normalsize
\vspace{.5cm}
Center for Particle Astrophysics, University of California,
Berkeley, CA 94720\\ 
and\\ 
National Energy Research Scientific Computing Center, Lawrence
Berkeley National Laboratory, University of California, Berkeley, CA
94720\\
\vspace{.5cm}
\end{center}
\baselineskip=24pt
\begin{abstract}
As the Cosmic Microwave Background (CMB) radiation is observed to
higher and higher angular resolution the size of the resulting
datasets becomes a serious constraint on their analysis. In particular
current algorithms to determine the location of, and curvature at, the
peak of the power spectrum likelihood function from a general
$N_{p}$-pixel CMB sky map scale as $O(N_{p}^{3})$. Moreover the
current best algorithm --- the quadratic estimator --- is a
Newton-Raphson iterative scheme and so requires a `sufficiently good'
starting point to guarantee convergence to the true maximum. Here we
present an algorithm to calculate bounds on the likelihood function at
any point in parameter space using Gaussian quadrature and show that,
judiciously applied, it scales as only $O(N_{p}^{7/3})$. Although it
provides no direct curvature information we show how this approach is
well-suited both to estimating cosmological parameters directly and to
providing a coarse map of the power spectrum likelihood function from
which to select the starting point for more refined techniques.
\end{abstract}
\begin{center}
{\small PACS numbers: 98.80, 98.70.V, 02.70}\\
\end{center}
\end{titlepage}
 
 
\section{Introduction}

Planned observations of the Cosmic Microwave Background (CMB) will
have sufficient angular resolution to probe the CMB power spectrum up
to multipoles $l \sim 1000$ or more (for a general review of
forthcoming observations see \cite{S}). If we are able to extract the
multipole amplitudes $C_{l}$ from the data sufficiently accurately we
will be able to obtain the values of the fundamental cosmological
parameters to unprecedented accuracy. The CMB will then have lived up
to its promise of being the most powerful discriminant between
cosmological models \cite{Kn,HSS,KKJS}.

Extracting the power spectrum is conceptually simple --- the raw data
is cleaned and converted into a time-ordered dataset. This is then
converted to a sky temperature map, which in turn is analysed to find
the location of, and curvature at, the maximum of the likelihood
function of the power spectrum. In practice as the size of the dataset
increases the problem rapidly becomes intractable by conventional
methods. This is particularly true of the final step --- the
likelihood analysis of any reasonably general sky temperature map.

An observation of the CMB contains both signal and noise
\begin{equation}
\Delta_{i} = s_{i} + n_{i}
\end{equation}
at each of $N_{p}$ pixels. For independent, zero-mean, signal and
noise the covariance matrix of the data
\begin{equation}
M \equiv \left< {\bf \Delta} \, {\bf \Delta}^{T} \right> 
= \left< {\bf s} \, {\bf s}^{T} \right> + 
  \left< {\bf n} \, {\bf n}^{T} \right> \equiv S + N
\end{equation}
is symmetric, positive definite and dense. For any theoretical power
spectrum $C_{l}$ we can construct the corresponding signal covariance
matrix $S(C_{l})$; knowing the noise covariance matrix $N$ for the
experiment we now know the observation covariance matrix for that
power spectrum $M(C_{l})$. The probability of the observation given
the assumed power spectrum is then
\begin{equation}
\label{eq.lf}
P({\bf \Delta} \, | \, C_{l}) = \frac{e^{-{\small\frac{1}{2}}
\, {\bf \Delta}^{T} M^{-1} {\bf \Delta}}} {(2 \pi)^{N_{p}/2} \left| M
\right|^{1/2}}
\end{equation}
Assuming a uniform prior, so that
\begin{equation}
P(C_{l} \, | \, {\bf \Delta}) \propto P({\bf \Delta} \, | \, C_{l})
\end{equation}
we can restrict our attention to evaluating the right hand side of
equation (\ref{eq.lf}). Unfortunately unless the noise covariance
matrix is unrealistically simple (eg. diagonal) both direct evaluation
\cite{G1,G2} and quadratic estimation \cite{T,BJK} of the likelihood
function scale as at best $O(N_{p}^{3})$ \cite{B}, making them
impractical for the forthcoming $10^4$ -- $10^6$ pixel datasets.

\section{Bounding The Likelihood Function}

Instead of the expensive exact evaluation of the likelihood function,
here we implement a much cheaper bounding algorithm due to Golub et al
\cite{GM,GS}. This method determines bounds
\begin{equation}
{\cal L} \leq {\bf u}^{T} f(A) \, {\bf u} \leq {\cal U}
\end{equation}
for an $n$-vector ${\bf u}$, symmetric positive definite $n \times n$
matrix $A$ and smooth function $f$ defined on the spectrum of $A$. The
underlying idea is to rewrite the problem as a Riemann-Stieltjes
integral which is then approximated using Gaussian quadrature. Since
$A$ is symmetric it can be written in eigendecomposition as
\begin{equation}
A = Q^{T} \Lambda \, Q
\end{equation}
where $Q$ is the orthogonal matrix of normalised eigenvectors and
$\Lambda$ is the diagonal matrix of increasing eigenvalues
$\Lambda_{ii} = \lambda_{i}$. Writing $\tilde{{\bf u}} \equiv Q \, {\bf
u}$ we have
\begin{eqnarray}
{\bf u}^{T} f(A) \, {\bf u} 
& = & {\bf \tilde{u}}^{T} f(\Lambda) \, {\bf \tilde{u}} \nonumber \\
& = & \sum_{i=1}^{n} f(\lambda_{i}) \, \tilde{u}_{i}^{2} \nonumber \\
& = & \int_{\lambda_{1}}^{\lambda_{n}} f(\lambda) \, {\rm d}\mu(\lambda) 
\equiv I(f)
\end{eqnarray}
where the measure $\mu(\lambda)$ is a piecewise constant function
defined by
\begin{equation}
\mu(\lambda) = \left\{
\begin{array}{ll}
0 & \;\;\;\; \lambda < \lambda_{1} \\
\sum_{i=1}^{j} \tilde{u}_{i}^{2} & \;\;\;\; \lambda_{j} \leq \lambda < \lambda_{j+1} \\
\sum_{i=1}^{n} \tilde{u}_{i}^{2} & \;\;\;\; \lambda_{n} \leq \lambda \\
\end{array}
\right.
\end{equation}
The integral $I(f)$ can now be bounded above and below using
Gauss-Radau quadrature \cite{DR}
\begin{eqnarray}
I(f) & = & \sum_{i=1}^{m} \omega_{i} f(\theta_{i}) + \nu_{1} f(\lambda_{1}) + R_{1} \nonumber \\
I(f) & = & \sum_{i=1}^{m} \omega_{i} f(\theta_{i}) + \nu_{n} f(\lambda_{n}) + R_{n} 
\end{eqnarray}
with weights $\omega_{i}$ and $\nu_{1,n}$, nodes $\theta_{i}$ and
$\lambda_{1,n}$, with opposite-signed remainders $R_{1,n}$.  Given the
resulting bounds on $I(f)$ we can increase the number of nodes $m$
until some convergence criterion, such as a maximum relative error,
\begin{equation}
\frac{{\cal U} - {\cal L}}{{\cal U} + {\cal L}} < \epsilon
\end{equation}
is met. Calculating such bounds scales as $O(m n^2)$ due to the
$O(n^2)$ matrix-vector multiplication in using the Lanczos algorithm
\cite{DR} to calculate each of the $m$ nodes and their weights.

Rewriting the likelihood function of equation (\ref{eq.lf}) as
\begin{equation}
P({\bf \Delta} \, | \, C_{l}) \propto \exp \left( - \frac{1}{2} \left(
{\bf \Delta}^{T} M^{-1} {\bf \Delta} + {\rm Tr} \left[ \ln M \right] \right) \right)
\end{equation}
leaves ${\bf \Delta}^{T} M^{-1} {\bf \Delta}$ and ${\rm Tr} \left[ \ln
M \right]$ to be evaluated. The former is already in the required form
and can be bounded immediately. For the latter we note that
\begin{equation}
\left< {\bf v}^{T} f(A) \, {\bf v} \right> = {\rm Tr} \left[ f(A) \right]
\end{equation}
where ${\bf v}$ is a random vector whose elements are $\pm 1$ with
equal probability. Generating $r$ realisations of ${\bf v}$ we can
calculate the bounds 
\begin{equation}
{\cal L}_{i} \leq {\bf v}_{i}^{T} \, \ln M \, {\bf v}_{i}
\leq {\cal U}_{i} \;\;\;\;\;\;\;\; 1 \leq i \leq r
\end{equation}
for each, from which we want to derive bounds on the expectation value
\begin{equation}
\mu \equiv \left< {\bf v}^{T} \, \ln M \, {\bf v} \right> = {\rm Tr} \left[ \ln M \right]
\end{equation}
For each realisation the estimator
\begin{equation}
X_{i} = \frac{1}{2} \left( {\cal U}_{i} + {\cal L}_{i} \right) = \mu + \delta_{i} + \epsilon_{i}
\end{equation}
the sum of the expectation value $\mu$, a random sample error $\delta_{i}$, 
and a systematic bound-width error $\epsilon_{i}$, where
\begin{equation}
\left| \epsilon_{i} \right| \leq \frac{1}{2} \left( {\cal U}_{i} - {\cal L}_{i} \right) \equiv a_{i}
\end{equation}
(note that this is an absolute, not relative, error measure). Given
the variance of the systematic-free data
\begin{eqnarray}
S^{2} & = & \frac{1}{r-1} \sum_{i=1}^{r} \left( \mu + \delta_{i} -
\overline{\mu + \delta_{i}} \right)^{2} \nonumber \\
      & = & \frac{1}{r-1} \sum_{i=1}^{r} \left( \delta_{i} - \bar{\delta} \right)^{2}
\end{eqnarray}
and assuming that $r$ is large enough for the central limit theorem to
apply, our estimator $\bar{X}$ has student's t-distribution with
$(r-1)$ degrees of freedom, and we can take bounds
\begin{equation}
\bar{X} - \frac{1}{\sqrt{r}} \, \tau_{r-1, \alpha} \, S - \bar{a} \leq \mu \leq
\bar{X} + \frac{1}{\sqrt{r}} \, \tau_{r-1, \alpha} \, S + \bar{a} 
\end{equation}
with $\alpha$ confidence. Although the systematics prevent us from
determining $S$ itself we can calculate a stochastically larger
quantity as follows: taking the sample variance of the midpoints
\begin{eqnarray}
S^{2}(X) & = & \frac{1}{r-1} \sum_{i=1}^{r} \left( \mu + \delta_{i} + \epsilon_{i} -
               \mu - \bar{\delta} - \bar{\epsilon} \right)^{2} \nonumber \\
	 & = & S^{2} + \frac{1}{r-1} \sum_{i=1}^{r} \left( 
		2 \, \left(\delta_{i} - \bar{\delta} \right) \, 
		     \left( \epsilon_{i} - \bar{\epsilon} \right) +
		\left( \epsilon_{i} - \bar{\epsilon} \right)^{2} \right) \nonumber \\
	 & \geq & S^2 - 2 \, S \, \sqrt{ \frac{1}{r-1} \sum_{i=1}^{r} a_{i}^{2}}
\end{eqnarray}
where we have used the fact that
\begin{equation}
\sum_{i=1}^{r} \left( \epsilon_{i} - \bar{\epsilon} \right)^{2} \leq
\sum_{i=1}^{r} \left( \epsilon_{i} \right)^{2} \leq
\sum_{i=1}^{r} a_{i}^{2}
\end{equation}
Thus, defining
\begin{equation}
A = \sqrt{ \frac{1}{r-1} \sum_{i=1}^{r} a_{i}^{2}}
\end{equation}
we have
\begin{equation}
S^{2} - 2 \, A \, S - S^{2}(X) \leq 0
\end{equation}
and hence the bound
\begin{equation}
S \leq A + \sqrt{ A^{2} + S^{2}(X) }
\end{equation}

If the number of terms $m$ required to evaluate each of the $r$ bound
pairs is approximately constant, we have a method to bound the
likelihood function which scales as $O(r m n^{2})$ overall.

\section{Results}

The viability of this approach depends crucially on the way that the
number of nodes in the Gauss-Radau quadrature ($m$) and the number of
trace estimates ($r$) depend on the size of the dataset ($n$) and the
required tightness of the bounds ($\epsilon$). To examine these
dependencies we have applied the algorithm to subsets of the COBE
data and a standard CDM target power spectrum.

The number of nodes required to achieve a given accuracy clearly
depends on that accuracy. At the extrema 
\begin{eqnarray}
\epsilon = 1 & \rightarrow & m = 1 \nonumber \\
\epsilon = 0 & \rightarrow & m = n
\end{eqnarray}
Figure 1 shows the dependence of the number of nodes on size of the
dataset for typical useful values $\epsilon = 10^{-1}$, $10^{-2}$,
$10^{-3}$, $10^{-4}$, and $10^{-5}$. The points are from numerical
experiments, evaluating bounds on ${\bf v}^{T} \ln M \, {\bf v}$ and
averaging over 100 realisations of ${\bf v}$. The solid lines are
power law fits $m \propto n^{\beta}$ (giving overall scaling as
$O(n^{2+\beta})$) with
\begin{equation}
\beta = \left\{ 
\begin{array}{lcr}
1/3 & {\rm for} & 10^{-1} \geq \epsilon \geq 10^{-2} \\
1/2 & {\rm for} & 10^{-3} \geq \epsilon \geq 10^{-5}
\end{array}
\right.
\end{equation}

Figure 2 shows the normalised 99\% confidence bounds achieved on ${\rm
Tr} \left[ \ln M \right]$ for a 1000 pixel dataset as the number of
estimates $r$ increases, with covergence set at $\epsilon = 10^{-1}$,
$10^{-2}$, $10^{-3}$, $10^{-4}$ and $10^{-5}$. Not surprisingly, the
10\% bounds on each estimate give a rather poor overall
constraint. However with the 1\% and tighter bounds we can determine
the logarithm of the likelihood function to 2 -- 3\% with 99\%
confidence in as few as 20 realisations. Note that below the 1\% level
the bounds are dominated by sample error, and become essentially
independent of the bound width. Figure 3 shows the 99\% confidence
bounds achieved after 20 realisations at $\epsilon = 10^{-2}$, showing
no systematic variation as the size of the dataset increases.

\section{Conclusions}

We have presented an algorithm to calculate probabilistic bounds on
the power spectrum likelihood function from an $N_{p}$-pixel CMB map
using Gaussian quadrature which scales as between $O(N_{p}^{7/3})$ and
$O(N_{p}^{5/2})$ --- a very significant advance on existing algorithms
for the exact calculation which scale as $O(N_{p}^{3})$. Since
lowering the convergence constraint below the 1\% level gains us only
marginally tighter final bounds at the expense of increasing the
scaling power, it is not recommended for the forthcoming $10^{4}$ --
$10^{6}$ pixel CMB maps. Our final algorithm of choice therefore gives
better than 3\% bounds on the logarithm of the likelihood function
with $O(N_{p}^{7/3})$ operations with 99\% confidence.

Since this algorithm gives no information about the local curvature of
the likelihood function it is not as well suited as quadratic
estimator techniques for searching a large multi-dimensional parameter
space for its likelihood maximum. However for direct estimation of a
small set of cosmological parameters this technique is certainly
viable and fast. Moreover, even when the parameters are taken to be
the multipole moments (individually or in bins), quadratic estimation,
being a Newton-Raphson iteration, requires a starting point
`sufficiently close' to the maximum to guarantee convergence; the
algorithm presented here is then well suited to provide a coarse
overall mapping of the likelihood function from which to select a
starting point for more refined techniques.

\section*{Acknowledgments}
 
This work was supported by the Laboratory Directed Research and
Development Program of Lawrence Berkeley National Laboratory under the
U.S. Department of Energy, Contract No. DE-ACO3-76SF00098, and used
resources of the National Energy Research Scientific Computing Center,
which is supported by the Office of Energy Research of the
U.S. Department of Energy. This work is also part of the COMBAT
project supported by NASA AISRP grant NAG5-3941. The author wishes to
thank Jim Demmel and Phil Stark for many helpful discussions.

\newpage
\nonfrenchspacing
\noindent
{\Large Figure Captions}
\vspace*{0.4in}

\noindent
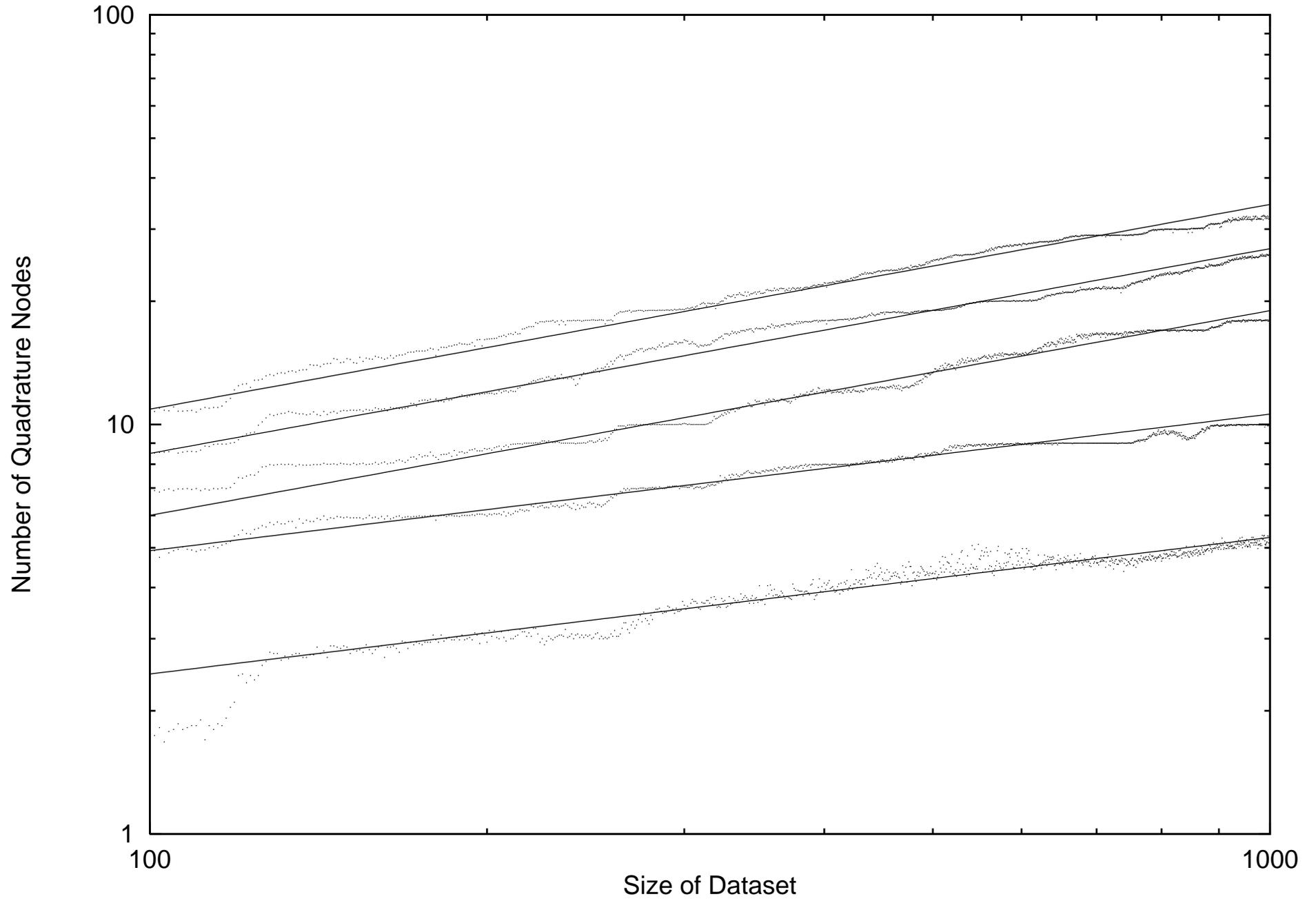
{\em Figure 1}\\ 
A log-log plot of the scaling in the number of quadrature nodes $m$
required to achieve bounds with a given relative error convergence
criterion $\epsilon$ with the size of the CMB dataset $n$. The points
are obtained from numerical experiments; the lines are power law fits
$m \propto n^{\beta}$. From the bottom of the figure reading upwards
the relative errors are $\epsilon = 10^{-1}, 10^{-2}, 10^{-3},
10^{-4}$ and $ 10^{-5}$, with corresponding power laws $\beta = 1/3,
1/3, 1/2, 1/2$ and $1/2$.
\vspace*{0.4in}

\noindent
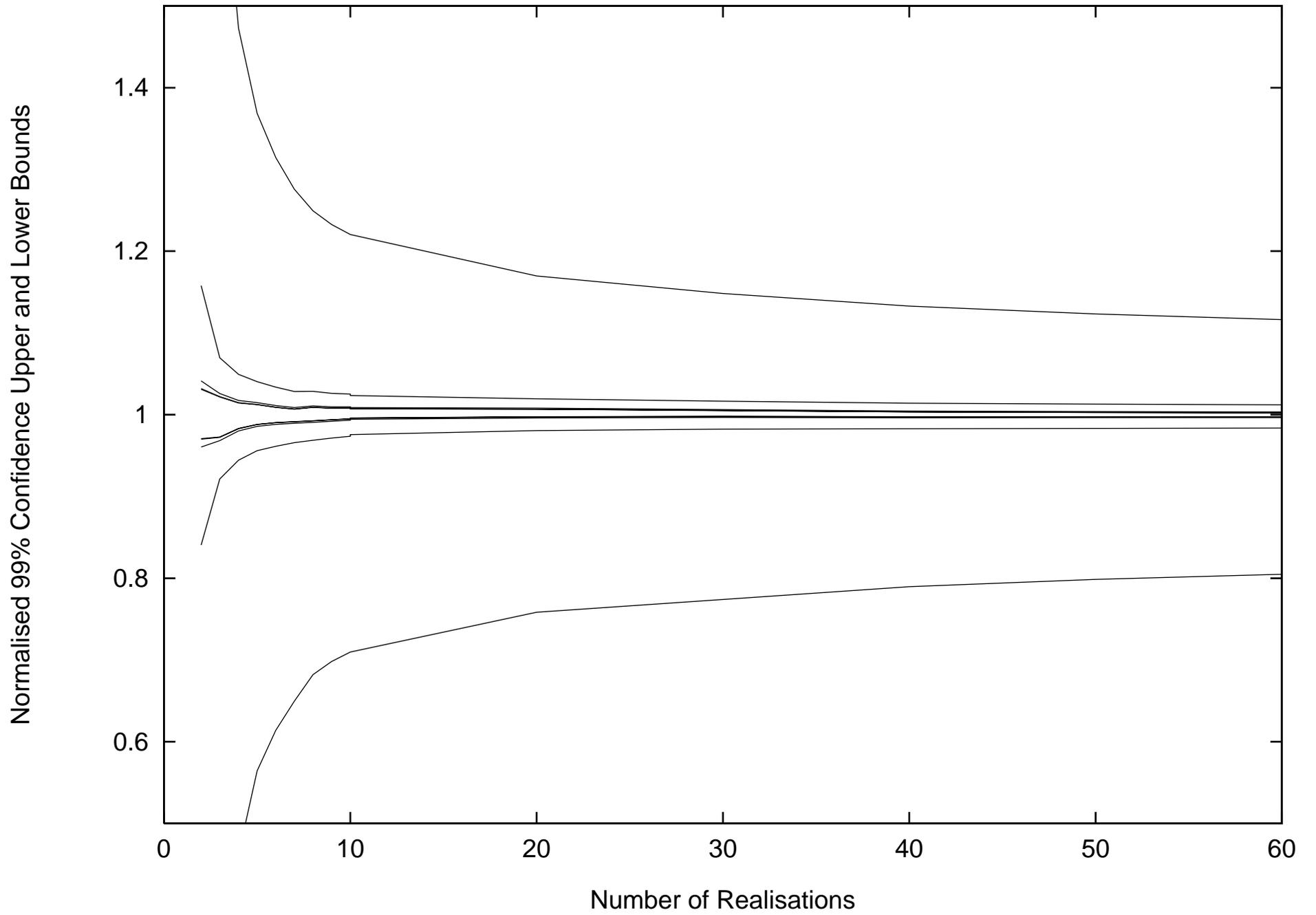
{\em Figure 2}\\ 
A plot of the normalised 99\% confidence upper and lower bounds
achieved on ${\rm Tr} \left[ \ln M \right]$ for a 1000 pixel dataset
against the estimator sample size $r$. From the outer limits reading
inwards the line-pairs correspond to the relative error covergence
criterion being set at $\epsilon = 10^{-1}, 10^{-2}, 10^{-3}, 10^{-4}$
and $10^{-5}$.
\vspace*{0.4in}

\noindent
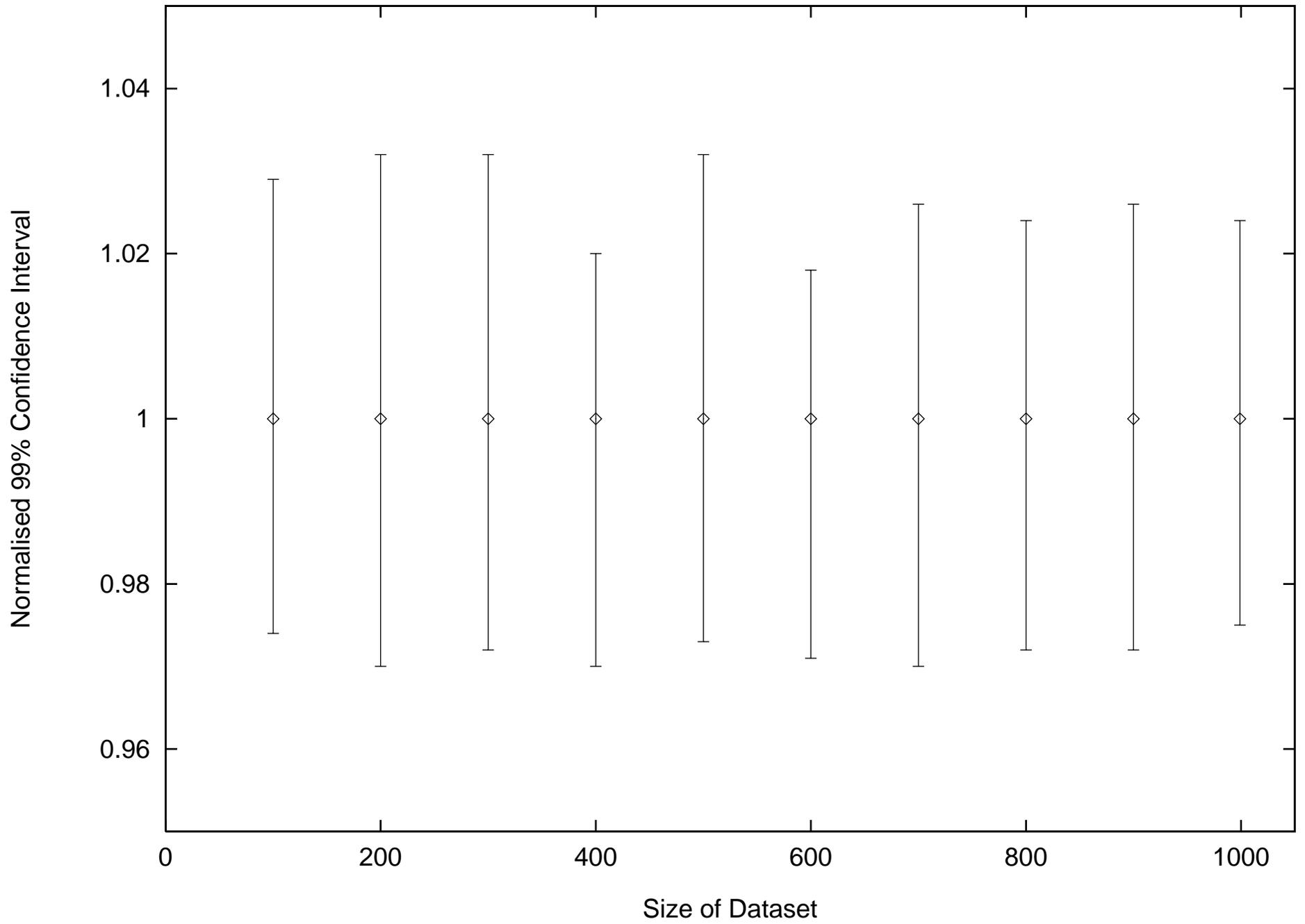
{\em Figure 3}\\
A plot of the variation in the normalised 99\% confidence upper and
lower bounds achieved on ${\rm Tr} \left[ \ln M \right]$ with the
relative error convergence criterion set at $\epsilon = 10^{-2}$ after
$r = 20$ estimates as the size of the CMB dataset $n$ increases.
\end{document}